%% file: main.tex
\documentclass[runningheads,anonymous=true]{llncs}
\usepackage[T1]{fontenc}
\usepackage{graphicx}
\usepackage{multirow}
\usepackage{graphicx}
\usepackage{subfigure}
\usepackage{threeparttable}
\usepackage{hyperref}
\usepackage{url}
\usepackage{color}

\usepackage{enumitem}
\usepackage{tablefootnote}
\usepackage[skip=5pt]{caption}
\usepackage{booktabs}
\usepackage{newtxmath}
\DeclareMathAlphabet {\mathbf}{OT1}{cmr}{bx}{n}

\def\covidia{Covidia}
\def\covidialow{covidia}

\begin{document}
\begin{sloppypar}

\pagestyle{headings}
\setcounter{page}{1}
\pagenumbering{arabic}
\title{\covidia: COVID-19 Interdisciplinary Academic Knowledge Graph}
%
%
\author{Cheng Deng\inst{1} \and Jiaxin Ding\inst{1} \and Luoyi Fu\inst{1} \and Weinan Zhang \inst{1} \and
Xinbing Wang\inst{1} \and Chenghu Zhou\inst{1}}
\institute{Shanghai Jiao Tong University
\\ \email{\{davendw, yiluofu\}@sjtu.edu.cn}}


%
\maketitle
\begin{abstract}

The pandemic of COVID-19 has inspired extensive works across different research fields. 
Existing literature and knowledge platforms on COVID-19 only focus on collecting papers on biology and medicine, neglecting the interdisciplinary efforts, which hurdles  knowledge sharing and research collaborations between fields to address the problem. 
Studying interdisciplinary researches requires effective paper category classification and efficient cross-domain knowledge extraction and integration. 
In this work, we propose \textbf{\covidia}, \textbf{COVID}-19 \textbf{i}nterdisciplinary \textbf{a}cademic knowledge graph to  bridge the gap between knowledge of COVID-19 on different domains. 
We design frameworks based on contrastive learning for disciplinary classification, and propose a new academic knowledge graph scheme for entity extraction, relation classification and ontology management in accordance with interdisciplinary researches. Based on \covidia{}, we also establish knowledge discovery benchmarks for finding COVID-19 research communities and predicting potential links.

\end{abstract}

\input{intro}
\input{overview}
\input{construction}
\input{evaluation}

\input{application}

\input{related}

\section{Conclusion}

In this paper, we propose a novel COVID-19 interdisciplinary academic knowledge graph,  \textbf{\covidia{}}, which extracts knowledge from all COVID-19 related research papers published in the major venues across different disciplines.  
The framework can not only benefit the researchers on COVID-19, but also be leveraged to studying potential future pandemics. 
The techniques we propose to generate interdisciplinary knowledge graphs are not limited to applications on COVID-19, but can also be applied to any scenarios where knowledge from different domains needs integrating. Finally, the entire system of \covidia{} and its resources will be publicly accessible after the final draft.


%
%
%
\bibliographystyle{splncs04}
\bibliography{ref}

\end{sloppypar}
\end{document}

%% file: intro.tex
\section{Introduction}


The pandemic of COVID-19 has aroused extensive attention of the academic researchers all around the world, huge volume of research works are conducted and published.   
Due to the profound and complex impact of COVID-19,  
the research works do not only include scientific discoveries in biology and medicine~\cite{huang2020clinical}, but also require the collaborations from other fields, such as computer science, sociology, mathematics, politics, etc.~\cite{garcia2020management,wang2020covid,de2022fully} to solve this problem with interdisciplinary insights~\cite{Kim2021ImpactsOC}. 
However, the boom of COVID-19 publications and the highly interdisciplinary collaborations
result in a ``paperdemic''~\cite{Yang2022PaperdemicDT,DinisOliveira2020COVID19RP}, which makes qualified information and knowledge retrieval harder. 
Organizing interdisciplinary works and extracting knowledge  is  a critical problem to solve. 

However, organizing interdisciplinary research efforts on COVID-19 is not paid enough attention to, which hurdles information and knowledge sharing between different fields. 
Existing COVID-19 literature datasets focus on collecting papers on biology and medicine~\cite{chen2021litcovid,trewartha2020covidscholar,Wang2020CORD19TC} without considering publications in other fields~\cite{dastani2021overview}.  
According to our statistics shown in \autoref{tab:dcdis}, 48\% out of 1.6 million publications on COVID-19 are not published in the venues of biology or medicine. 
Besides, even though part of interdisciplinary works are published in the biology and medicine venues, these works are not properly classified into all other disciplines they belong to~\cite{Tanaka2018RecategorizingIA}, which makes it hard for researchers to draw the interdisciplinary insights, and difficult 
for interdisciplinary researchers to find exact matches of their interests. 
Organizing interdisciplinary works is not simply putting together all papers related to COVID-19, but organizing works into the exact categories labelled by all related disciplines.   


\begin{figure}[t]
    \centering
    \includegraphics[width=\linewidth]{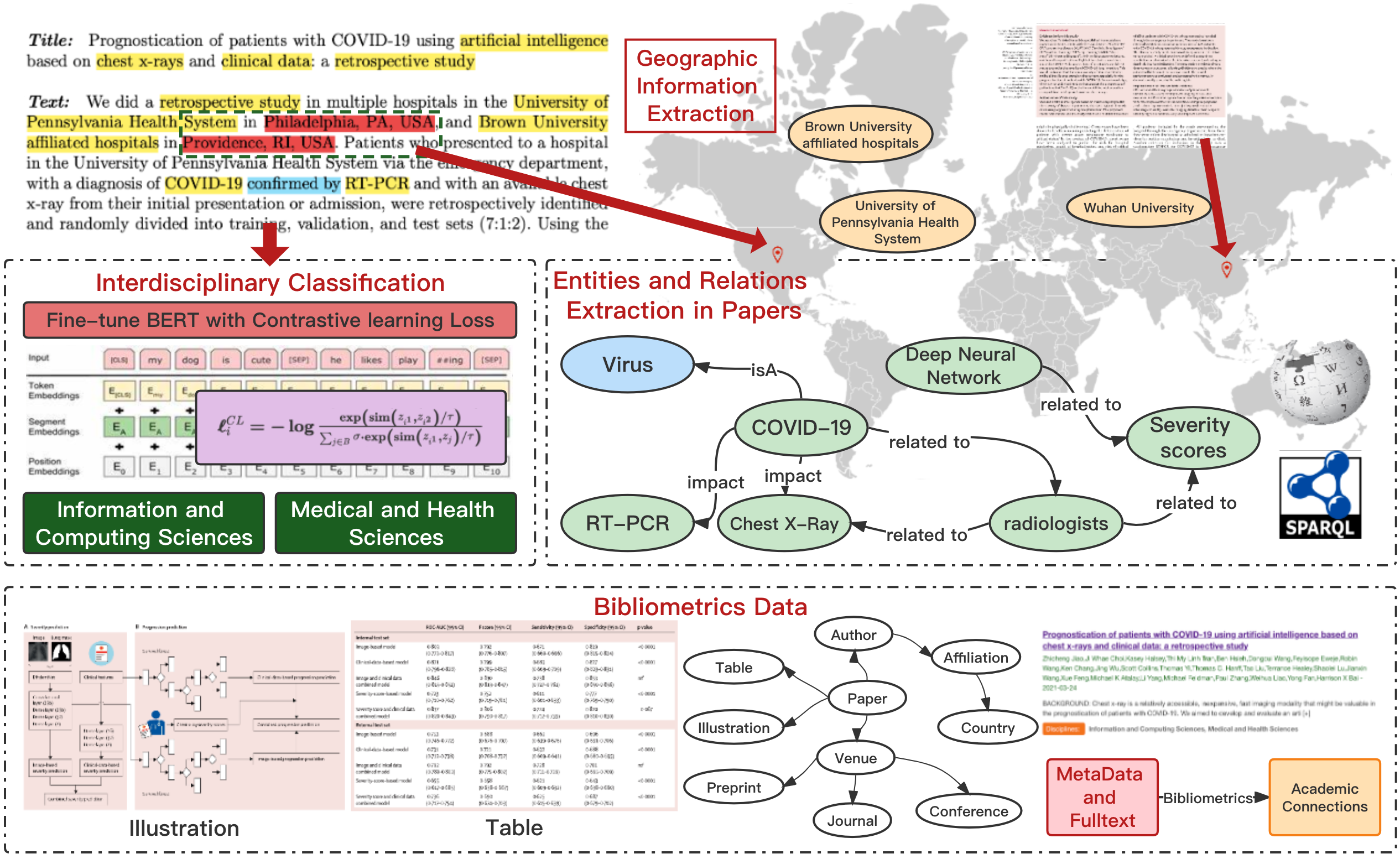}
    \caption{\covidia{} Overview: From Paper to Knowledge.}
    \label{covidia-overview}
    \vspace{-1em}
\end{figure}

Further, knowledge extraction in interdisciplinary researches is not well studied, which makes knowledge isolated in respective fields. For example, COVID-KG~\cite{wang2020covid} extracts the bio-pharmaceutical and protein entities, and establishes relations between entities from papers on COVID-19 in biomedical fields, 
while GAKG~\cite{deng2021gakg} mines the knowledge entities in geoscience area. 
The ways of extracting knowledge entities on these different disciplines are different, 
the knowledge entities varies from each other, 
and even same words in different fields can have different meanings.  
Therefore, it is challenging to extract knowledge from interdisciplinary works and integrate respective knowledge extracted. 


In face of the above challenges, we propose frameworks to solve interdisciplinary paper classification and knowledge extraction. 
First, we propose a multi-label paper classification model with contrastive learning on different disciplines. 
Thereafter, we enhance the entity extraction model to interdisciplinary scenarios by aligning the entities to contents in Wikipedia to effectively extract entities with disciplinary contexts and efficiently adapt our model to open domains of interdisciplinary researches.  
We also propose an academic knowledge scheme graph, where knowledge entities on ontology layers can be connected by 
sharing same papers or bibliometric entities on instance layers, and vise versa, 
for interdisciplinary knowledge integration. 

In this work, we collect interdisciplinary papers on COVID-19, with size of 1.5 million, and propose \covidia{}, a \textbf{COVID}-19 \textbf{i}nterdisciplinary \textbf{a}cademic knowledge graph (KG), to address the problems of information retrieval and knowledge extraction of interdisciplinary researches on COVID-19. 
An overview of \covidia{} with examples is demonstrated in \autoref{covidia-overview}, the entire system and resources will be publicly accessible after the final draft, and our contributions can be summarized as follows:

\begin{itemize}[leftmargin=0.8em]
  \item[1.] In this paper, we propose an ongoing interdisciplinary academic knowledge graph on COVID-19,  \textbf{\covidia{}}, which summarizes bibliometrics data, domain glossaries, illustrations, tables, and spatio-temporal  information of papers and relations between them. To our knowledge, \covidia{} is the largest knowledge graph and academic research literature  platform for COVID-19.
  \item[2.] This paper achieves interdisciplinary paper classification by introducing contrastive learning over interdisciplinary categories.  
We introduce an entity extraction model based on  learning to rank and a BERT-based relation extraction model using segment embedding. We deploy this model and extract knowledge entities in the COVID-19 articles and relations between these entities. 
  \item[3.] In \covidia{}, we put forward a new academic knowledge graph scheme, adding disciplinary categorization to knowledge entities, which facilitates researchers on information retrieval and mining.
\end{itemize}

%% file: overview.tex
\vspace{-2em}
\begin{figure}[h]
    \centering
    \includegraphics[width=\linewidth]{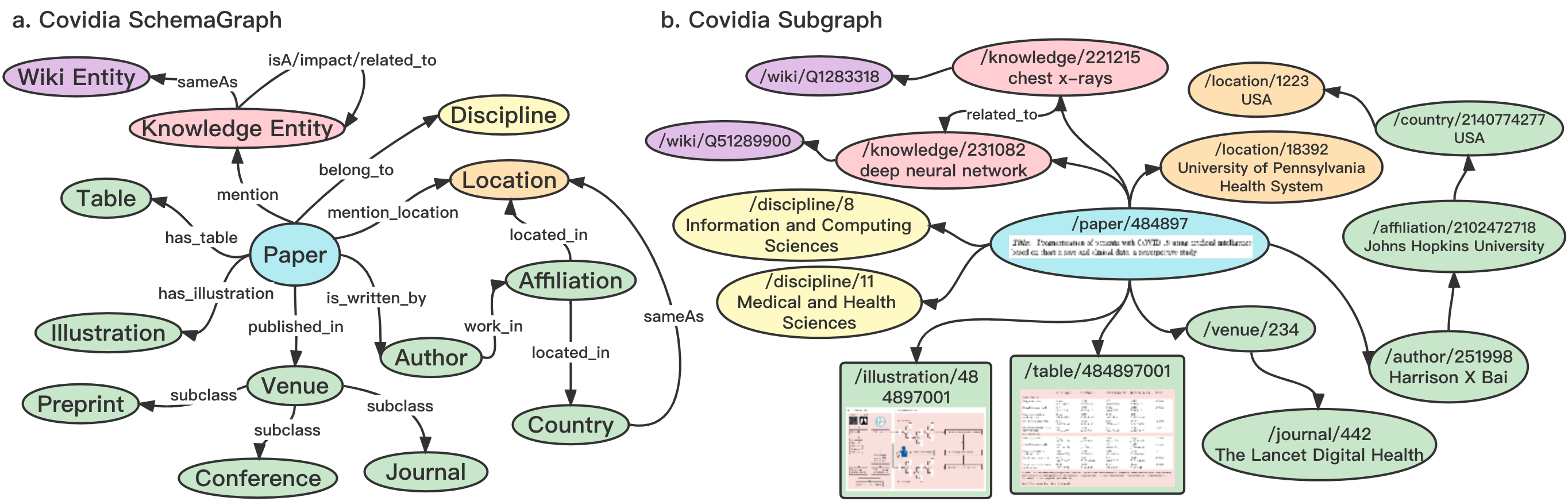}
    \caption{\covidia{} Schema,  Figure a. is the schema-graph of \covidia{} and Figure b. is an example and a subgraph of \covidia{}}
    \label{kgsg}
    \vspace{-3em}
\end{figure}

\section{\covidia{} Overview}
We first present the overview of \covidia{}. 
There are \textbf{13} concepts connected by \textbf{15} relations in the \covidia{} and \textbf{34} data properties. The distribution of the disciplines, are shown in \autoref{tab:dcdis}. 
We can see that about half of the publications on COVID-19 are not published in the venues of biology or medicine science. 
To obtain fine-grained knowledge in COVID-19 papers, we design the concept \textit{\covidialow:knowledge} representing the knowledge entities extracted from the papers. 
Meanwhile, \covidia{} use relation \textit{\covidialow r:mention\_knowledge} to connect the knowledge entities and papers, design relations \textit{\covidialow r:is\_A}, \textit{\covidialow r:impact} and \textit{\covidialow r:related\_to}, then adopt \textit{owl:sameAs} axioms linking knowledge entities to the outer entities. Furthermore, \textit{geohash} is used to represented the points of interest (POI).
The schema-graph of the \covidia{} is shown in \autoref{kgsg}. 

The statistics of \covidia{} concepts and relations is shown in \autoref{tab:statistic}. 
The knowledge graph consists of 
\textbf{6,097,866} instances and \textbf{24,705,417} links. 
To facilitate scholars in data mining, knowledge engineering, and information retrieval to browse and access \covidia{} data efficiently, we provide a SPARQL endpoint. 

\vspace{-2em}
\begin{table}[]
\caption{Statistics Concepts and Relations in \covidia{}}
\label{tab:statistic}
\resizebox{\linewidth}{!}{%
\begin{tabular}{@{}llll|llll@{}}
\toprule
\textbf{Concepts} & \textbf{Count} & \textbf{Concepts} & \textbf{Count} & \textbf{Relations} & \textbf{Count} & \textbf{Relations} & \textbf{Count} \\ \midrule
paper & 1,631,519 & topic & 17,334 & is\_cited\_by & 3,970,632 & work\_in & 2,370,233 \\
author & 3,087,916 & discipline & 22 & is\_written\_by & 8,433,200 & is\_located\_in & 314,158 \\
organization & 19,194 & papertable & 843,418 & is\_published\_in & 1,615,358 & has\_papertable & 643,418 \\
journal & 69,364 & illustration & 1,217,539 & in\_the\_topic\_of & 1,945,205 & has\_illustration & 817,539 \\
conference & 9,205 & knowledge & 1,563,932 & belongs\_to & 2,903,771 & isA/impact/related\_to & 3,829,201 \\
preprint & 58 & location & 68,465 & mention\_knowledge & 3,228,392 & rdfs:subClassOf & 45,231 \\
venue & 78,627 & geohash & 68,465 & mention\_location & 1,250,738 & owl:sameAs & 794,187 \\ \midrule
\multicolumn{2}{c}{\textbf{Total Entities}} & \multicolumn{2}{c|}{\textbf{8,675,058}} & \multicolumn{2}{c}{\textbf{Total Relations}} & \multicolumn{2}{c}{\textbf{32,161,263}} \\ \bottomrule
\end{tabular}%
}
\end{table}
\vspace{-2em}
\begin{table}[]
\centering
\caption{Disciplines Distribution in Covidia.}
\label{tab:dcdis}
\resizebox{\linewidth}{!}{%
\begin{tabular}{@{}cc|cc@{}}
\toprule
\textbf{Discipline} & \textbf{Count} & \textbf{Discipline} & \textbf{Count} \\ \midrule
Mathematical   Sciences & 206,651 & Medical and Health Sciences & 862,080 \\
Physical Sciences & 103,523 & Built Environment and Design & 29,997 \\
Chemical Sciences & 119,490 & Education & 92,892 \\
Earth Sciences & 17,198 & Economics & 56,758 \\
Environmental Sciences & 85,103 & Commerce, Management, Tourism and   Services & 75,453 \\
Biological Sciences & 304,991 & Studies in Human Society & 195,012 \\
Agricultural and Veterinary Sciences & 33,647 & Psychology and Cognitive Sciences & 118,497 \\
Information and Computing Sciences & 145,152 & Law and Legal Studies & 26,991 \\
Engineering & 234,416 & Studies in Creative Arts and Writing & 11,311 \\
Technology & 21,642 & Language, Communication and Culture & 78,571 \\
Medical and Health Sciences & 862,080 & History and Archaeology & 48,190 \\
Built Environment and Design & 29,997 & Philosophy and Religious Studies & 36,206 \\ \bottomrule
\end{tabular}%
}
\end{table}
\vspace{-2em}

%% file: construction.tex
\section{\covidia{} Construction}
In this section, we detail processes of building \covidia{}. We first obtain data integrated from different paper sources  and perform multi-label document classification to classify the papers. Meanwhile, we tag the specialized terms mentioned in the articles. We perform open-domain information extraction for each COVID-19-related paper, align the entity to the glossary, and classify their relations.

\subsection{Collection and Fusion of Bibliometrics Data}

We fuse the data by integrating papers from AceMap \cite{tan2016acemap}, CORD-19 \cite{Wang2020CORD19TC}, Digital Science \cite{porter2020covid}, and 58 preprint sites that were flagged as being related to COVID-19, normalizing institutions, and naming scholars. The details of papers, authors, organizations and venues for each data source are listed in \autoref{tab:covidsource}.

\begin{table}[h]
    \centering
    \caption{Statistics of the \covidia{} Data Sources}
    \label{tab:covidsource}
    \begin{tabular}{@{}lcccc@{}}
    \toprule
                          & Paper       & Author    & Organization  & Venue     \\ \midrule
    Acemap                & 204,940     & 725,525   & 10,161        & 14,438    \\
    CORD-19               & 640,833     & 1,425,106 & 80,121        & 119,166   \\
    Digital Science       & 693,919     & 2,088,412 & 1,367,993     & 37,311     \\
    Preprint Website      & 44,355      & 197,901   & 76            & 58        \\ \midrule
    \textbf{After Fusion} & 1,107,008   & 2,636,699 & 19,194        & 78,627    \\ \bottomrule
    \end{tabular}
\end{table}

We extract the papers marked as \textit{``2019 20 coronavirus outbreak''}, \textit{``severe acute respiratory syndrome coronavirus 2''} and \textit{``Coronavirus disease 2019''} in Acemap, collect the papers published by CORD-19 since January 2020, and collected the COVID-19 papers collected by Digital Science, using the query shared \cite{porter2020covid}, with the type of \textit{``article''}, \textit{``proceeding''} and \textit{``preprint''}. At last, we collect papers that study COVID-19 from 50 mainstream preprint websites. Each data source focuses on a different application scenario. 
We adopt scholars' name disambiguation and organization normalization algorithms~\cite{tan2016acemap} to finish the information fusion.
We build a BERT-based named entity recognition model and use SpaCy \cite{spacy} tools to extract the geographical and political locations. After strict normalization rules of the location entities, we deploy a GeoCoder to get the location coordinates. Finally, we use the Geohash algorithm to store coordinates to present the distribution of COVID-19 papers on a world map online.

All above, \covidia{} gathers all papers, scholars, institutions, and related bibliometric information about COVID-19 with papers' illustrations, tables, knowledge instances, and geographic locations extracted from the text. 

\subsection{Classification of Interdisciplinary Papers}

 Pre-trained language models based document classification model are the most widely used \cite{gutierrez2020document}. Similar to \cite{sun2019fine}, we develop an interdisciplinary classification model that combines the embeddings generated by SciBERT and contrastive learning loss with  \textbf{220,330} interdisciplinary annotated labels from Acemap. 


To perform single-task fine-tune BERT, we first choose the \textit{BCE Loss} with logits viewing multi-disciplinary classification as multiple binary classification,
\begin{equation}
   \ell_{k}^{BCE} = y_{i} \log(x_{i}) + (1-y_{i}) \log(1-x_{i}),
\end{equation}
where $k$ is denoted as the index of the batch, for each sample with index $i$, $x_{i}$ is the predicted label, $y_{i}$ is the ground true label. Meanwhile, intuitively two articles with totally different labels should be placed with a longer distance in the latent space. Similar to \cite{van2018representation,zhang2021supporting}, \covidia{} adopts a comparative learning loss function \textit{InfoNCE Loss}, for each document $z$ we generate a pair of augmentations for each sample in a batch $B$ (with a size $|B|$), to reduce the loss as: 
\begin{equation}
\ell_{i}^{C L}=-\log \frac{\exp \left(sim\left(z_{i^{1}}, z_{i^{2}}\right) / \tau\right)}{\sum_{j \in B} \sigma \cdot \exp \left(sim\left(z_{i^{1}}, z_{j}\right) / \tau\right)},
\end{equation}
where $z_{i^{1}}, z_{i^{2}}$ is a positive pair, $\sigma$ is an indicator function and $\tau$ denotes the temperature parameter setting as 0.5. The final loss function is:
\begin{equation}
    Loss = \ell_{k}^{BCE} + \ell_{k}^{CL}.
\end{equation}

We employ this model as a baseline, discussed in section 4. After multi-label discipline classification, we get the discipline distribution as shown in \autoref{mmmcd}:

\begin{figure}[]
    \centering
    \includegraphics[width=\linewidth]{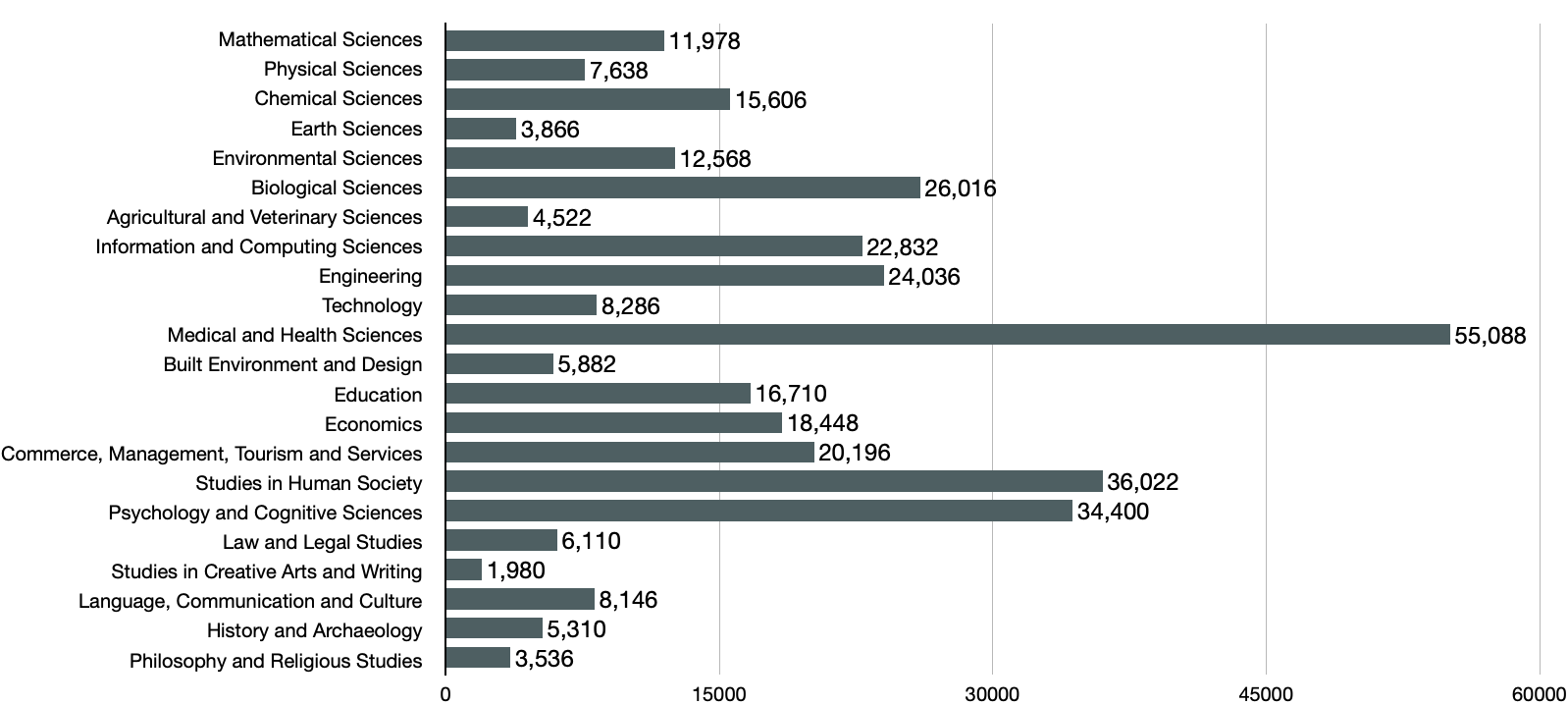}
    \caption{Disciplines Distribution in \covidia{}.} 
    \label{mmmcd}
\end{figure}

At last, we use this fine-tuned model as a new encoder to generate all the embeddings of the paper in \covidia{}.

\subsection{Extraction of Knowledge Entities}

Distribution of discipline glossary  can provide potential connections between papers that have neither citation relations nor the same authors.  Now we introduce the pipeline to extract knowledge instances and the referred locations from papers text.

For 22 disciplines, we have collected high-quality wiki entities, corresponding discipline glossary sets and discipline knowledge graphs as a set of disciplines knowledge. We ensure that each discipline has a high-quality discipline glossary, with disciplines knowledge graphs as replenishment. Meanwhile, for the sake of disambiguation, we invite experts from various disciplines to separate entities and claim them as the concept \textit{\covidia{}:knowledge}, so that they can be linked to the entities in the original knowledge graphs through the assertion of the \textit{\#sameAs}.

With the above disciplines knowledge sets, discipline knowledge graphs and wiki entities, we need to match the sentences in the papers semantically. Once the sentences in the text are semantically close to or directly refer to the discipline glossary sets and discipline knowledge entities, we can claim the relation of \textit{mention\_knowledge}.
Therefore, referring to explicit semantic analysis (ESA) \cite{egozi2011concept}, we first convert the terms that may be associated in an abstract through the vector transformation of TF-IDF~\cite{Salton1988TermWeightingAI}. In this process, we regard all the glossary and nodes in knowledge graphs in the table as entities, and their descriptions as documents $D$, view papers' abstracts as a query $q$ to find the words in the text. Thus, we can get the candidate entities $E$.

\begin{equation}
    E=Q(q, D), \quad E=\left\{e_{i}\right\}, i \geq 0 .
\end{equation}

Second, we take each entity's TF-IDF score, length, complexity, and letters' amount as the feature vector, consulting the LambdaRank \cite{quoc2007learning}, one of the learning to rank algorithms, we try to learn the function where given a text $q$, return with $n$-dims entities $E$, with related $n$-dims scores $S$, like $f(q, E)=S, \quad E=e_{i}, S=s_{i}, i \geq 0 ,$. Then we train a two-layer neural network, combining the feature vectors of entities and a loss function towards a pair as equation 6 to ensure that factor NDCG \cite{jarvelin2017ir} can finally reach a certain level.

\begin{equation}
    Loss_{i j}=\log \left\{1+\exp \left(-\sigma\left(s_{i}-s_{j}\right)\right)\right\} \cdot\left|\Delta N D C G_{i j}\right|,
\end{equation}
where $\sigma$ is a parameter shaping the \textit{sigmoid} function. The binary label we used to calculate the NDCG is annotated by experts from various disciplines, according to every pair of paper and candidate entities generated by the ESA step. 

\begin{figure}[h]
    \centering
    \includegraphics[width=\linewidth]{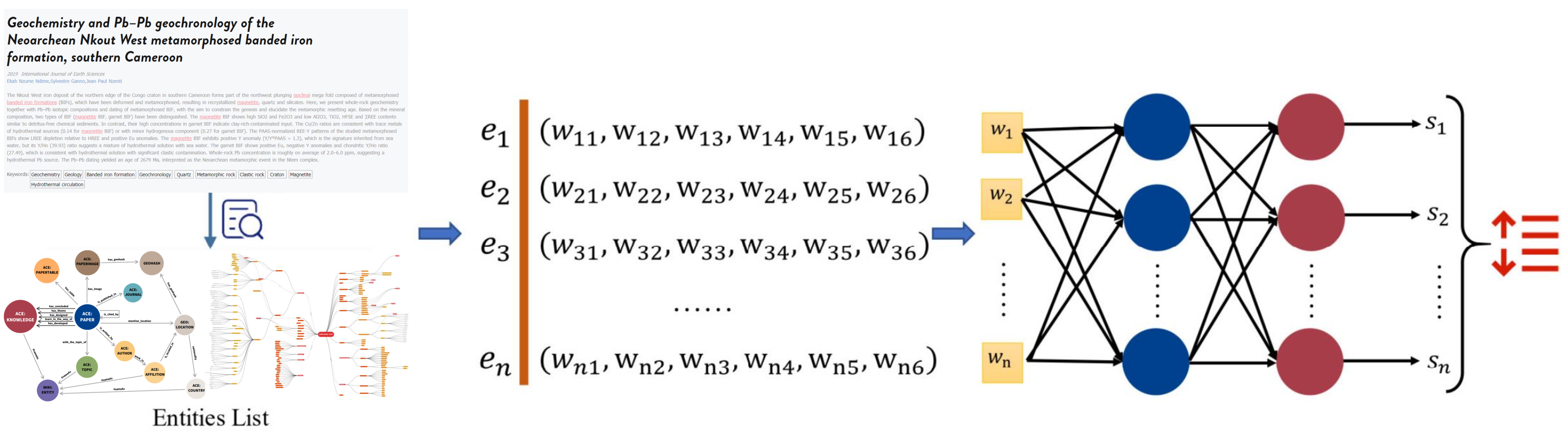}
    \caption{The Pipeline of the Ranking Model.}
    \label{l2r}
    \vspace{-2em}
\end{figure}

Finally, by setting the threshold, a considerable result is obtained. In this process, we sacrificed the recall rate in order to ensure accuracy. Overall, the precision of our model on the benchmark we set is \textbf{0.914}, and the recall rate is \textbf{0.391}. The pipeline of the model is shown in \autoref{l2r}.

\subsection{Classification of Relation Between Knowledge Entities}

In addition to bibliometric data, \covidia{} also mines the relations between knowledge entities mentioned in the papers to construct a knowledge graph for each paper. 
By recognizing the relations between knowledge entities of the papers, research can do reasoning over knowledge entities. Referring to the definitions of knowledge engineering and common sense knowledge graph, we define three relations \textit{is\_A}, \textit{impact}, and \textit{related\_to}. We first extract triples through the general open-domain triple extraction tool \textit{OpenIE} \cite{angeli2015leveraging}. For each paper, we extracted an average of \textbf{53} triples. For each paper $p$ we have
\begin{equation}
    OpenIE(p) = {(h_{o}, r, t _{o})},
\end{equation}
where $h_{o}$ and $t_{o}$ denotes the origin entities, and $r$ indicates the relations extracted by the \textit{OpenIE}. Then, we align the obtained head entity and tail entity with the entity obtained in the previous section by adopting the exact match rule to ensure accuracy to align and get the normalized head entity and tail entity $h$ and $t$. After that, we obtain persistent annotation data by deploying the Human-in-the-Loop system. We set the obtained annotations quadruple \textit{(sentence, head entity, relation, tail entity, label)} as $(s, g, r, t, l)$, where $l$ is \textit{\covidialow{}r:is\_A}, \textit{\covidialow{}r:has\_impact}, \textit{\covidialow{}r:is\_A}, and \textit{unknown}. 
\vspace{-1em}
\begin{figure}[h]
    \centering
    \includegraphics[width=0.8\linewidth]{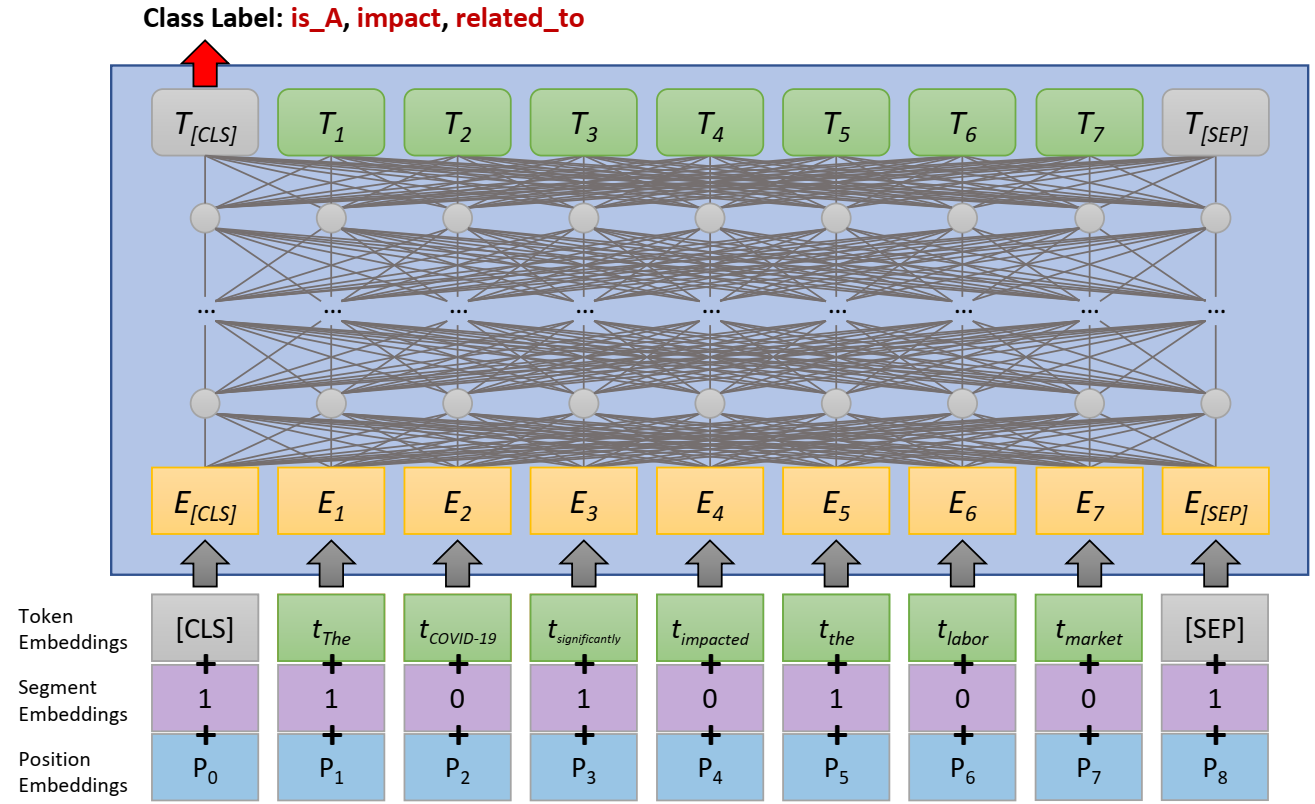}
    \caption{Input embedding of interdisciplinary classification model.}
    \label{openiebert}
    \vspace{-1em}
\end{figure}

Based on BERT, we design a linear layer as the task-specific layer for relation classification. The input to BERT is a single sentence with annotated entities and relations. Then we use the sum of three vectors to encode the input, of which token embedding is the WordPiece word vector, and segment embedding is specially designed, (embedding value 1 if the token is one of the elements in the triple, and the rest is 0), and a trainable position embedding. Finally, the probability is obtained through the $Softmax$ layer, and the output probability is as follows,
\begin{equation}
    p(l \mid \mathbf{h})=Softmax(\mathbf{W h}),
\end{equation}
where $\mathbf{h}$ stands for the output of the encoder and $\mathbf{W}$ is the parameter of the linear layer.
Furthermore, the model parameters are optimized by maximizing the log-probability of the correct label. The specific process is as \autoref{openiebert}. Finally, we process the model over the entire COVID-19 papers.

%% file: evaluation.tex
\section{Evaluation}

This section will evaluate and demonstrate the methods we use to construct the \covidia{}. We make benchmarks for Multi-Label Document Classification, phrase extraction, and open domain relation classification on our dataset and compare several baselines. In multi-disciplinary classification, we choose AUC as the measure, use precision and recall in the experiment for phrase extraction and open domain relationship classification, and obtain the final result by delineating a reasonable threshold. All the benchmarks can be accessed after the final draft.

\subsection{Interdisciplinary Papers Classification} 

\subsubsection{Dataset}

\begin{figure}[h]
    \centering
    \includegraphics[width=\linewidth]{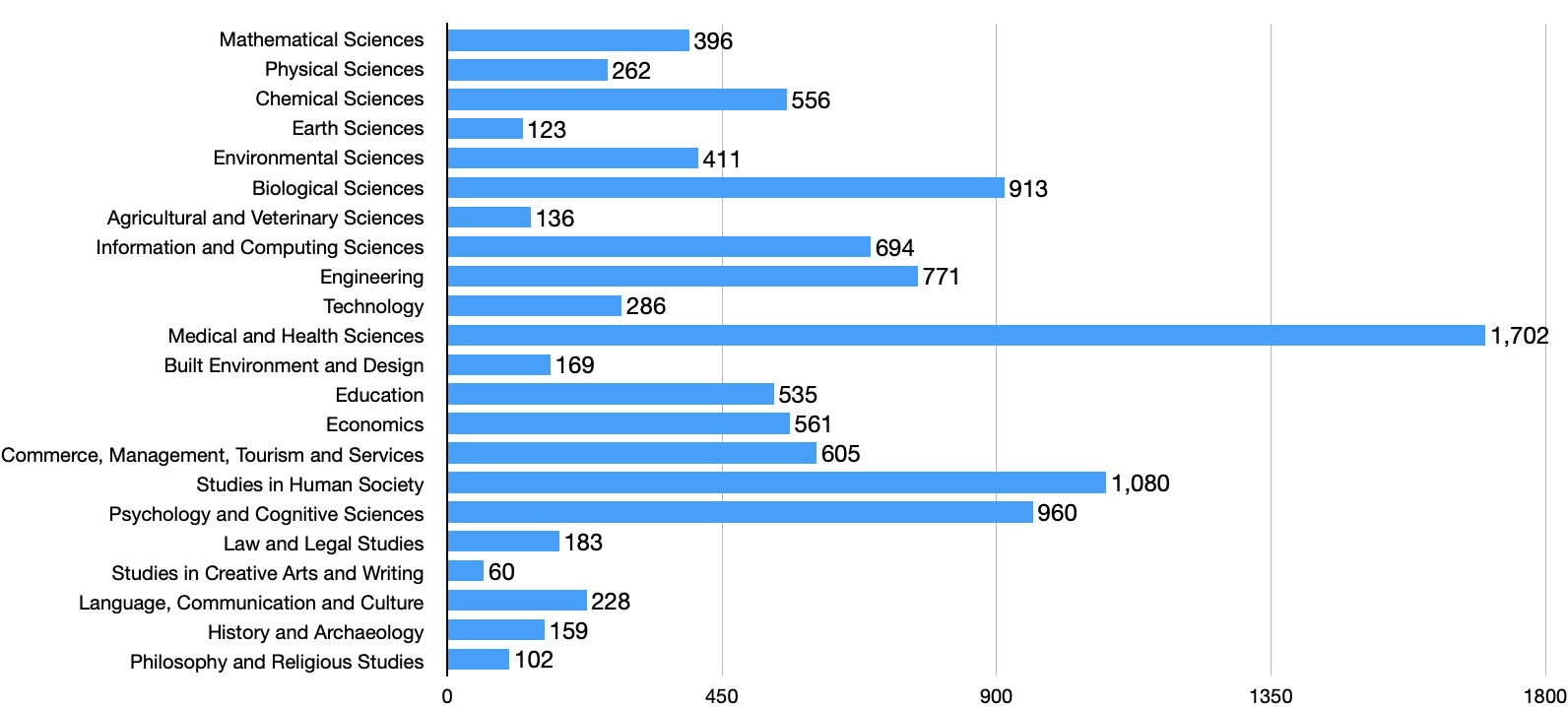}
    \caption{The distribution of the disciplines in the training set.}
    \label{cdcd}
    \vspace{-2em}
\end{figure}

We also constructed a mapping of COVID-19 articles and discipline labels based on the \covidia{}. Each article has at least two to four discipline labels. Our discipline labels are collected from digital science. The entire dataset contains 40,000 articles, including the articles' titles, abstracts, and IDs. For this dataset, the distribution of disciplines is as \autoref{cdcd}.
\vspace{-1.5em}

\subsubsection{Evaluation}

We assessed our method and some other baselines on the task of multi-label document classification. We compare LAHA \cite{huang2021label}, \textbf{SciBERT+FT}, and \textbf{SciBERT+CL} where SciBERT+FT is a model finetuning the SciBERT with BCE Loss on language model downstream multi-label classification task; SciBERT+CL is the model using contrastive learning mechanism.

As shown in the \autoref{tab:mldc_exp}, the model trained with contrastive learning loss has the highest AUC, although the precision and ranking metric NDCG is comparable. In order to enable the embedding of articles to contain classification information of different disciplines, we use the contrastive learning-based model.
\vspace{-1em}
\begin{table}[h]
    \centering
    \caption{Experimental results of Multi-label Document Classification}
    \label{tab:mldc_exp}
    \begin{tabular}{@{}llllll@{}}
        \toprule
        & Pre.@3 & Pre.@5 & NDCG@3 & NDCG@5 & AUC \\ \midrule
        LAHA & 88.34 & \textbf{77.78} & 91.31 & 89.30 & 95.12 \\
        SciBERT+FT & 87.99 & 76.58 & 91.18 & 88.48 & 95.24 \\
        SciBERT+CL & \textbf{89.20} & 77.61 & \textbf{92.17} & \textbf{90.32} & \textbf{96.33} \\ \bottomrule
    \end{tabular}
    \vspace{-1em}
\end{table}

\subsection{Knowledge Entities Extraction}

\subsubsection{Dataset}
By deploying an annotation system for Human-In-The-Loop machine learning, we allow experts in all disciplines to score our predicted phrases and build a ranking dataset to find the top entities in tagging phrases. The dataset has collected more than 1000 articles with 50,000 phrase annotations.
\vspace{-1em}
\begin{table}[h]
    \centering
    \caption{Experimental results of Phrase Tagging}
    \label{tab:pt_exp}
    \begin{tabular}{@{}lll@{}}
    \toprule
    Methods & Precision & Recall \\ \midrule
    ESA+BERT+L2R ($1^{st}$ Loop) & 0.859 & 0.292  \\
    ESA+BERT+L2R ($2^{nd}$ Loop) & 0.872 & 0.331 \\ 
    ESA+Word2Vec+L2R ($1^{st}$ Loop) & 0.841 & 0.319 \\
    ESA+Word2Vec+L2R ($2^{nd}$ Loop) & 0.854 & 0.347 \\ 
    ESA+TF-IDF+L2R ($1^{st}$ Loop) & 0.881 & 0.332 \\
    ESA+TF-IDF+L2R ($2^{nd}$ Loop) & \textbf{0.914} & \textbf{0.391} \\ 
    \bottomrule
    \end{tabular}
    \vspace{-2em}
\end{table}

\subsubsection{Evaluation}
For the phrase tagging task, we compare \textbf{ESA+L2R} model using TF-IDF, using BERT and Word2Vec as the feature constructing on our dataset. Since we have manual annotations to help improve the model's performance in cycle to guarantee the precision, we sacrifice the recall so that our model can tag out more accurate phrases. Meanwhile, we choose the best threshold on the test set and test the selection over the benchmark. The results are shown in \autoref{tab:pt_exp}, showing that the mechanism using TF-IDF in the second loop performs better than the other models.

\subsection{Relation Classification}

\subsubsection{Dataset}
Relation classification is the task of identifying the semantic relation between two entities in the text. With the help of the annotation system, we gather more than two thousand \textit{(triple, sentence, label)} records. For the open domain, we have designed two baselines for comparison to classify the relation in the open domain into three categories, including \textit{is\_A}, \textit{subclass\_of} and a superclass \textit{related\_to} to generalize the rest.
\vspace{-1em}
\begin{table}[h]
    \centering
    \caption{Experimental results of Open-domain Relation Classification}
    \label{tab:orc_exp}
    \begin{tabular}{@{}lll@{}}
    \toprule
     & Precision & Recall \\ \midrule
    SciBERT+FT & 76.53 & 71.18 \\
    SciBERT+SEG & \textbf{79.09} & \textbf{75.21} \\ \bottomrule
    \end{tabular}
    \vspace{-1em}
\end{table}
\subsubsection{Evaluation}

We set 400 of them as the benchmark to evaluate the models. We use precision and recall to evaluate these two models, and the results are in \autoref{tab:orc_exp}, indicating that with re-designed segment embedding, the pre-train model can perform better.

%% file: application.tex
\section{\covidia{} Application}

\covidia{} has been applied to two COVID-19 academic scenarios, including but not limited to retrieval of COVID-19-related papers via POI and semantic search over \covidia{} RDF dataset. Meanwhile, the \covidia{} can also be applied to network data mining as a heterogeneous network. 

\begin{figure}[h]
    \centering
    \subfigure[]{\label{mapa}
    \includegraphics[width=0.46\linewidth,height=0.30\linewidth]{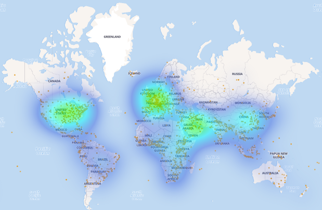}}
    \subfigure[]{\label{mapb}
    \includegraphics[width=0.46\linewidth,height=0.30\linewidth]{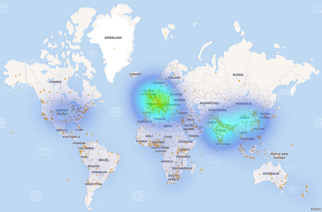}}
    \subfigure[]{\label{mapc}
    \includegraphics[width=0.46\linewidth,height=0.30\linewidth]{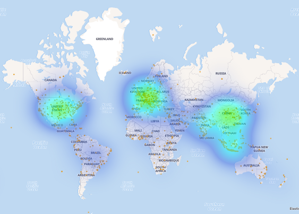}}
    \subfigure[]{\label{mapd}
    \includegraphics[width=0.46\linewidth,height=0.30\linewidth]{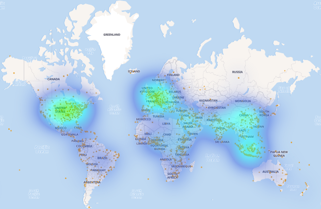}}
    \caption{Papers distribution on earth regarding different topics. (a) Keywords: ``Woman'', (b) Keywords: ``Lockdown'', (c) Discipline: ``Information and Computing Sciences'', (d) Discipline: ``Psychology and Cognitive Sciences''.}
    \label{map}
    \vspace{-2em}
\end{figure}

\subsection{Geographic Search over \covidia{}}

We provide the geographic locations mentioned in each paper. By visualizing the results, as shown in \autoref{map}, we can see that the distribution of searched papers varies for different keywords. When we set the keyword to \textit{``women''}, we can find that both Eurasia and North America are scorching areas, but when we look for papers about \textit{``lockdown''}, the difference between Eurasia and the Americas is apparent, which can be explained by different quarantine policies in different regions. Coincidentally, the discipline distribution of \textit{``computer science''} tends to be slightly less than that of \textit{``psychology''}. 

Based on geography information in \covidia{}, we provide a geographic search over \covidia{}. First, researchers can input any phrases to query in \covidia{}, including papers' titles, abstracts, disciplines, and knowledge entities the COVID-19 papers mentioned. Researchers enter keywords in the input box, e.g., ``Vaccine in China'', and the related papers would be shown on the map. Besides, for better user interaction, users can drag the map, zoom in and zoom out the map to search for papers within the bounding box with specified keywords. 

\subsection{Semantic Search over \covidia{}}

Using \covidia{}, researchers can learn more about the relationship in COVID-19 academia. \covidia{} perform several examples, including one-hop queries, such as returning articles that mention a particular knowledge point, two-hop queries, such as retrieving illustrations in a particular area and retrieving places that an affiliation usually studies as an example of a three-hop query. These queries can be used in scientific research and scholarly communication about COVID-19. Moreover, these queries are generally unanswerable by existing search engines and differ from the current academic platform.

\subsection{Network Science on \covidia{}}

\begin{table}[]
\vspace{-2em}
\caption{Statistics of \covidia{} Social Science Benchmarks}
\label{tab:socialnetwork}
\resizebox{\textwidth}{!}{%
\begin{threeparttable}
\begin{tabular}{@{}c|c|ccccccc@{}}
\toprule
\textbf{Networks} & \textbf{Concepts} & \textbf{Size} & \textbf{Volume} & \textbf{Max Degree} & \textbf{Avg Degree} & \textbf{$\alpha$} & \textbf{$p\_value$} & \textbf{$x\_min$} \\ \midrule
\textbf{Coauthor Network} & author & 303,995 & 4,853,932 & 1,286 & 15.96 & 1.436 & 0.613 & 246 \\
\textbf{Citation Network} & paper & 398,920 & 5,218,680 & 15,980 & 13.08 & 1.503 & 0.107 & 184 \\
\multirow{2}{*}{\begin{tabular}[c]{@{}c@{}}\textbf{Author-Paper Network} \\ (Author-writes-Paper)\end{tabular}} & author & 2,636,703 & \multirow{2}{*}{5,444,318} & 2,770 & 2.06 & 1.511 & 0.912 & 121 \\
 & paper & 967,070 &  & 5,368 & 5.63 & 1.32 & 0.748 & 104 \\
\multirow{2}{*}{\begin{tabular}[c]{@{}c@{}}\textbf{Paper-Author Network} \\ (Paper-inspires-Author)\end{tabular}} & author & 981,419 & \multirow{2}{*}{17,418,387} & 2,701 & 17.75 & 1.377 & 0.193 & 401 \\
 & paper & 247,338 &  & 99,836 & 70.42 & 1.569 & 0.737 & 596 \\ \bottomrule
\end{tabular}%
\begin{tablenotes}
    \footnotesize
    \item[1] $\alpha, p, x_{min}$ is used to measure the power-law distribution characters.
\end{tablenotes}
\end{threeparttable}
}
\end{table}

\begin{figure}[h]
    \centering
    \subfigure[Citation Network]{\label{disa}
    \includegraphics[width=0.48\linewidth,height=0.30\linewidth]{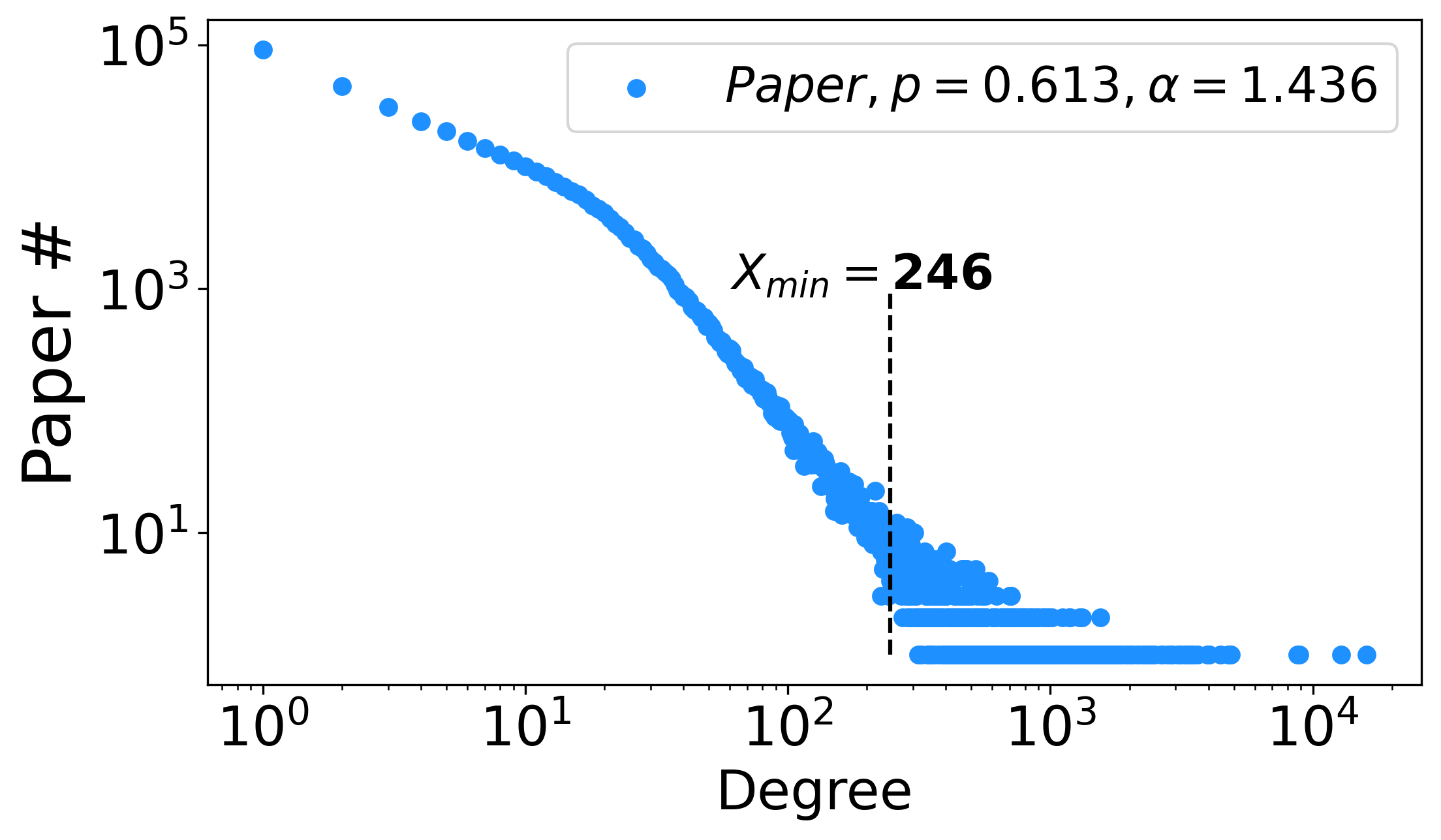}}
    \subfigure[Author-Paper Network]{\label{disb}
    \includegraphics[width=0.48\linewidth,height=0.30\linewidth]{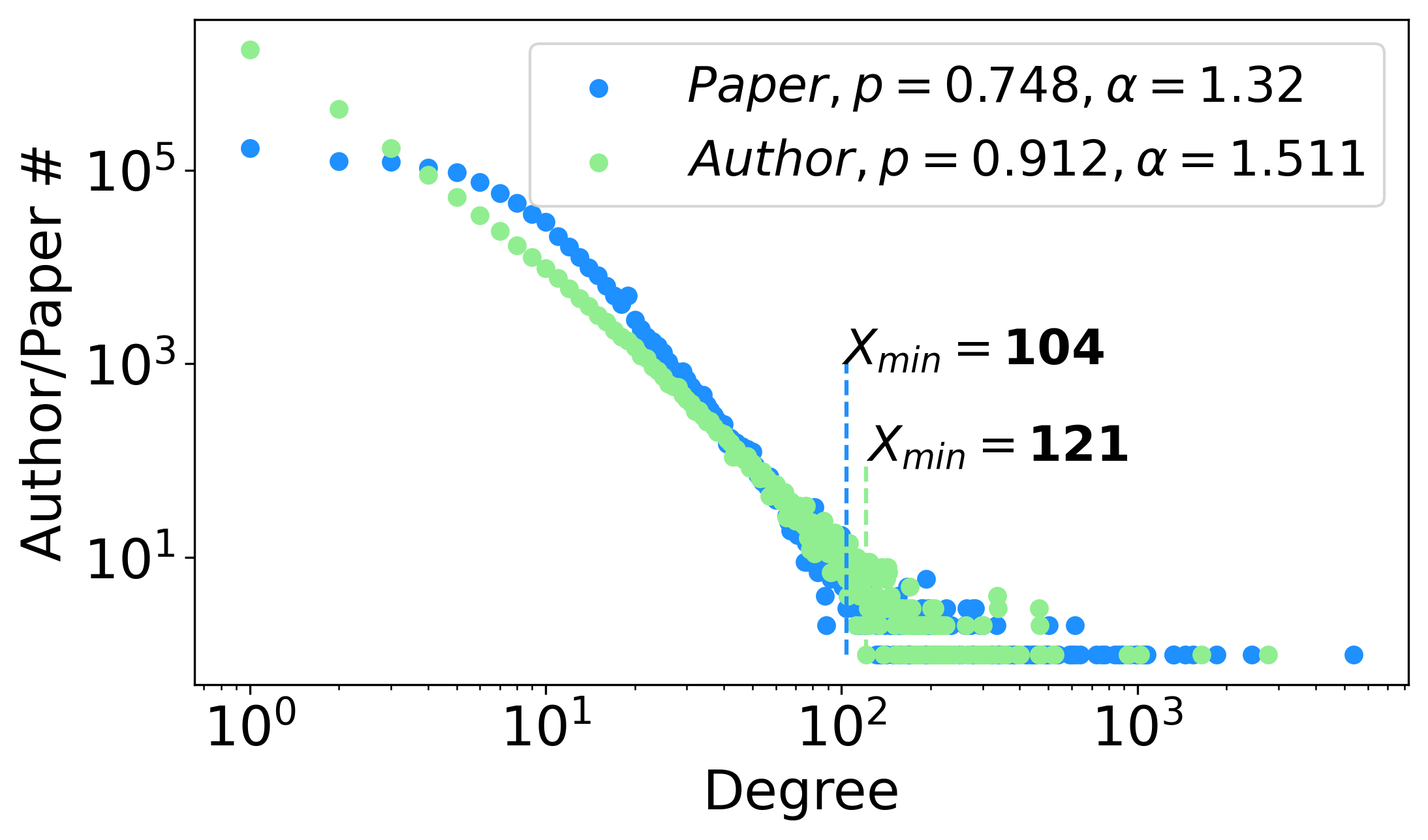}}
    \subfigure[Coauthor Network]{\label{disc}
    \includegraphics[width=0.48\linewidth,height=0.30\linewidth]{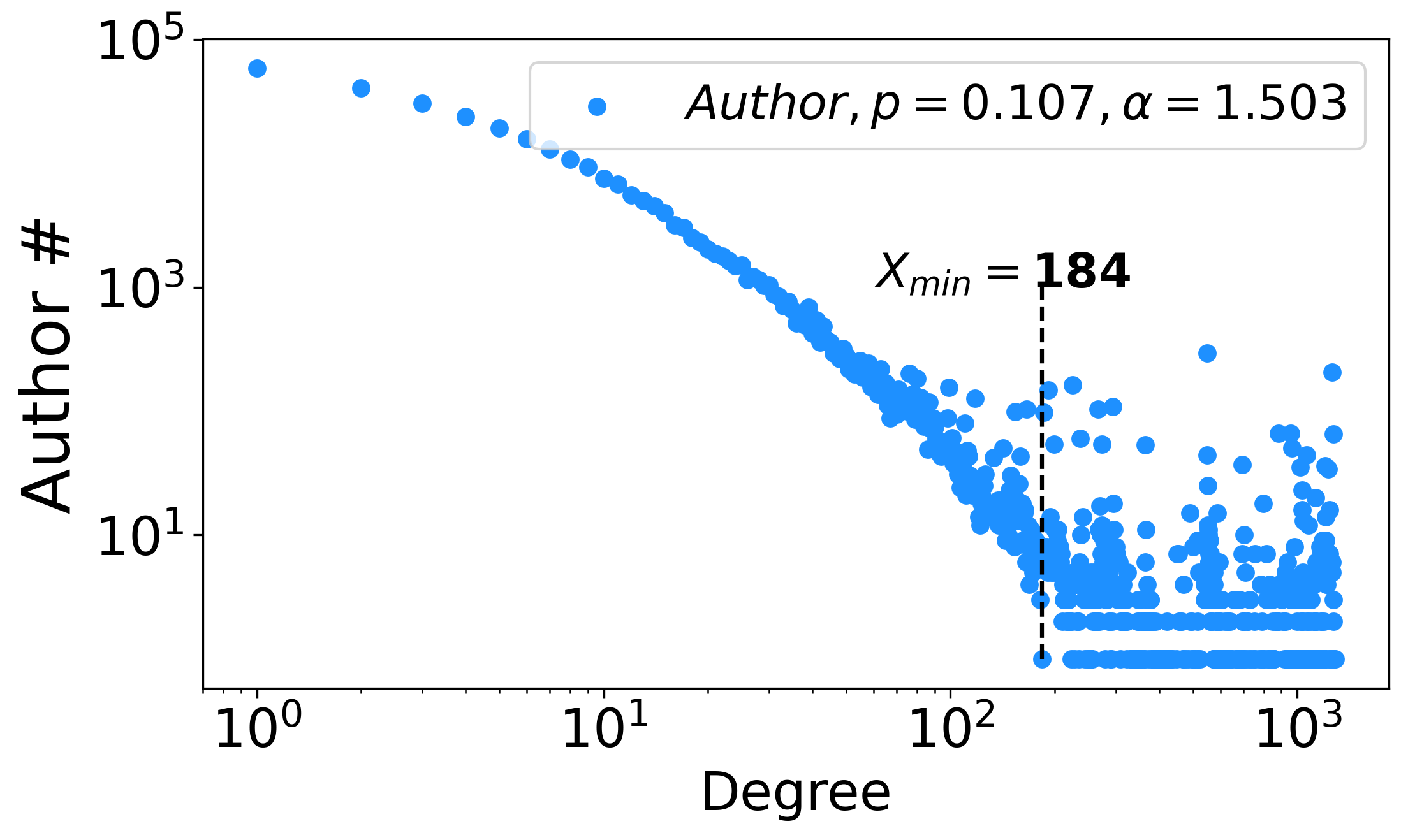}}
    \subfigure[Paper-Author Network]{\label{disd}
    \includegraphics[width=0.48\linewidth,height=0.30\linewidth]{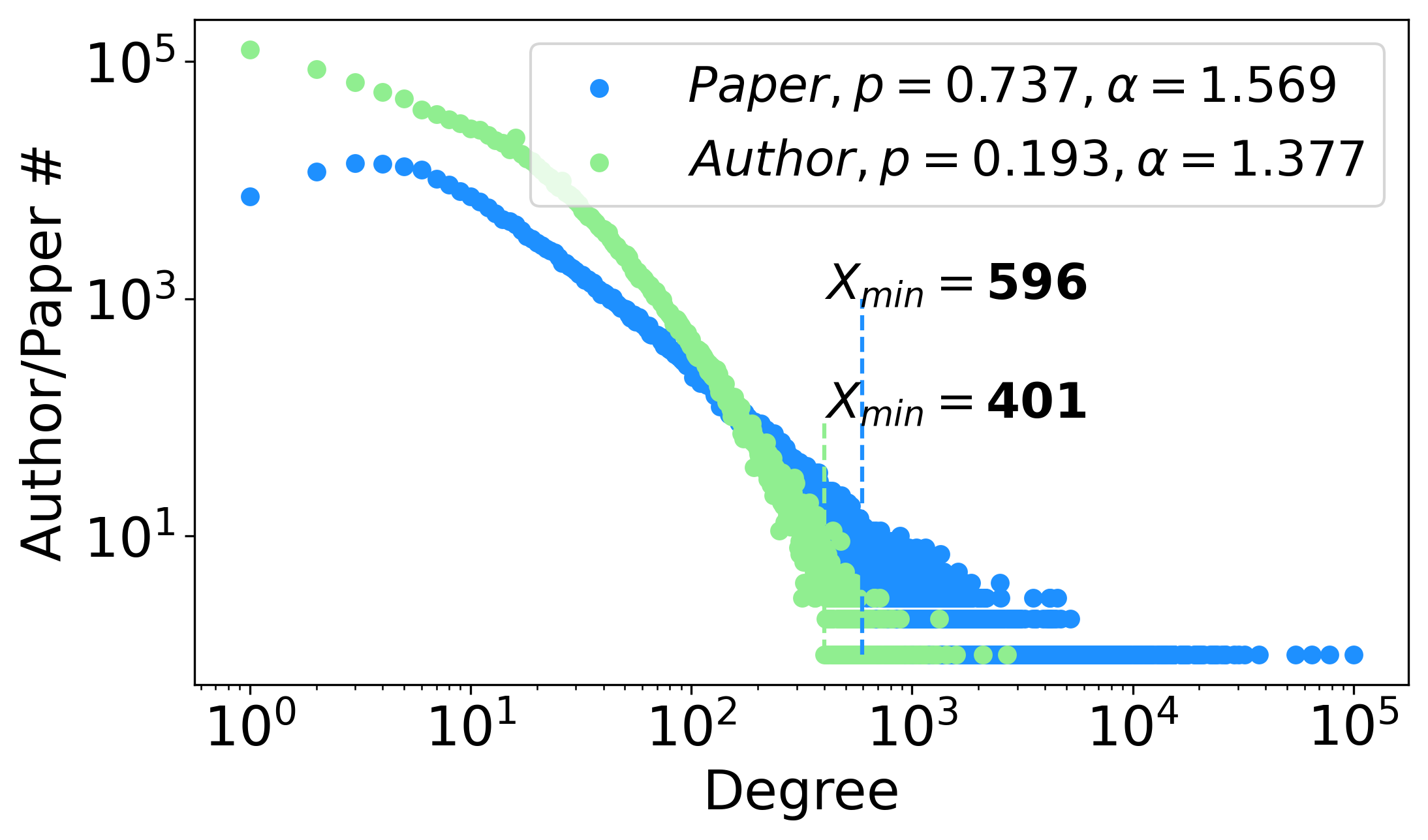}}
    \caption{Degree distribution on Network Science Benchmarks.}
    \label{dis}
\end{figure}

\covidia{} provides four social networks  to network science. There are two homogeneous networks, an author cooperation network and a citation network, which can be used for author and paper classification and COVID-19 community detection. Two paper-author networks are also obtained, including a bipartite graph network between articles written by authors and a bipartite graph network between authors and papers cited by their papers, which can be used for reference recommendations. The relevant data statistics and the corresponding distribution can be found in \autoref{tab:socialnetwork} and \autoref{dis}.

%% file: related.tex
\section{Related Work}

We briefly review the related work of \covidia{}, including literature classification, information extraction, and COVID-19 related knowledge graph.

\textbf{Literature Classification.} Since the outbreak of COVID-19, researchers have deployed natural language processing models and developed suitable methods to understand the pandemic-related text material~\cite{Chen2021ArtificialI}. \cite{gutierrez2020document} provides an analysis of several multi-label document classification models on the LitCOVID dataset, and pre-trained language models~\cite{sun2019fine} outperform others. \cite{pappagari2019hierarchical,huang2021label} use label correlation to estimate the similarity between papers. \cite{Wagner2011ApproachesTU,Tanaka2018RecategorizingIA} classify and evaluate interdisciplinary papers by designing catalogs and indicators.

\textbf{Information Extraction in Literature.} When building a KG, mining entities and relations are main challenges. In the task of entity extraction, BERT-based~\cite{devlin2018bert} and BiLSTM-CRF~\cite{huang2015bidirectional} are mainstream, which have been deployed in domain-specific scenarios like material science~\cite{weston2019named} and geoscience~\cite{deng2021gakg}. What's more, AutoPhrase~\cite{shang2018automated} sheds light on unsupervised phase tagging. For  relation extraction, the works of mining relations between entities extracted from literature are rare and mainly in biomedical articles~\cite{ashburner2000gene,chen2021litcovid,doctorevidence}. Meanwhile, the means like OpenRE~\cite{hanetal2019opennre} share the idea of distant supervision on close-domain relation extraction, while OpenIE~\cite{angeli2015leveraging} gives the open-domain task a brand new paradigm. 

\textbf{COVID-19 related Knowledge Graph.} In COVID-19 scenarios, dimensions.ai~\cite{porter2020covid} provides a subset of Digital Science via a set of keyword queries, while CORD-19~\cite{Wang2020CORD19TC} provides a machine-readable research dataset. COVID-KG~\cite{wang2020covid}, CKG~\cite{wise2020covid}, COVID-19 KG RDF database~\cite{Chen2021COVID19KG} and KG-COVID-19~\cite{reese2021kg} has collected the COVID-19 related literature metadata and some of the key terms related to the papers.
and cause-and-effect KG on COVID-19 pathophysiology is proposed by~\cite{domingo2021covid}. A framework that can integrate heterogeneous biomedical data to produce KGs was developed for COVID-19~\cite{reese2021kg}.